\documentclass{epl}

\title{ Monte Carlo study of Bose Laughlin wave function
        for filling factors $1/2$, $1/4$ and $1/6$ }

\author{ Orion Ciftja  }

\institute{
         Department of Physics, Prairie View A\&M University,
         Prairie View, Texas 77446, USA }


\begin{document}

\maketitle

\begin{abstract}
Strongly correlated two-dimensional electronic systems 
subject to a perpendicular magnetic field at 
lowest Landau level (LLL) filling factors: $1/2$, $1/4$ and $1/6$ 
are believed to be composite fermion (CF) Fermi liquid phases.
Even though a Bose Laughlin wave function cannot describe these
filling factors we investigate whether such a wave function provides 
a {\it lower energy bound} to the true CF Fermi liquid energies.
By using Monte Carlo simulations in disk geometry 
we compute the Bose Laughlin energies and compare them
to corresponding results for the spin-polarized LLL CF Fermi liquid 
state and avalable data from literature.
We find the unexpected result that, for filling factors $\nu=1/4$ and $1/6$, 
the Bose Laughlin ground state energy is practically identical to the 
true CF liquid energy while this is not the case at $\nu=1/2$ where
the Bose Laughlin ground state energy is sizeably lower than the energy 
of the CF Fermi liquid state.
\end{abstract}

Modulation doped $GaAs/Al_{x}Ga_{1-x}As$ heterojunctions 
provide an almost ideal experimental realization  of a
two-dimensional electron gas (2DEG).
In these devices, the application of a strong perpendicular magnetic 
field quenches the kinetic energy and eventually brings the electrons 
in a regime where correlations are of utmost importance.
In cases where the perpendicular magnetic field is very strong all electrons
partially occupy the lowest Landau level (LLL) and have their spins 
fully polarized in the direction of the magnetic field.
The phase diagram of a 2DEG in a strong perpendicular magnetic field
at filling factors $0 < \nu \leq 1$ is intricate with competing liquid and
Wigner solid phases.
At filling factors, $\nu=1/3, 2/5, 3/7, \ldots$ 
and $\nu=1/5, 2/9, \ldots$
electrons condense into an incompressible liquid state
and the fractional quantum Hall effect (FQHE)  occurs.~\cite{tsui} 
%
It is believed that at even denominator filling factors, $\nu=1/2, 1/4$
and $1/6$ the electrons form a compressible liquid 
state~\cite{willett1,willett2,willett3}
while for filling factors, $ 0 < \nu \leq 1/7$, Wigner 
crystallization occurs.~\cite{lam,esfarjani,zhu1,zhu2}

On the theoretical side, the accurate prediction of 
the critical filling factor, $\nu_c$ where the liquid-solid transition 
occurs is very difficult.
For an ideal 2DEG system free of any disorder the best available 
theoretical estimates~\cite{lam} suggest $\nu_c \cong 1/6.5$.
In experimental realm, the situation is more complex. 
Various measurements~\cite{reentrant} on high quality samples, 
yet with a finite amount of disorder, show the existence of a reentrant 
solidlike phase around the FQHE state, $\nu=1/5$ which likely represents
an electronic solid phase pinned to fluctuations of the potential.
The majority of theoretical studies which are based on models free of 
any disorder agree that the critical filling factor, $\nu_c$ 
below which the Wigner crystal forms is close to and probably 
slightly larger than
filling $1/7$ in excellent agreement with the estimate of 
Lam and Girvin.~\cite{lam}.
A recent study~\cite{kunyang} based on the exact diagonalization method
found strong evidence that the Wigner crystal forms at filling factors,
$\nu \leq 1/7$.
Such study considered the filling factors, $\nu=1/6, 1/7$, and $1/8$.
It was found that for $\nu=1/8$, as well as $\nu=1/7$ the translational
symmetry (signature of liquid phase) is broken and the system has 2D
crystalline order.
However, the system continued to have its translational symmetry for the
case of filling factor $\nu=1/6$, suggesting that the value of the 
critical filling factor, $\nu_c$ is between slightly above $1/7$, but
below $1/6$.

The principal FQHE states at $\nu=1/3$ and $1/5$ are thoroughly
explained and very well described by the Laughin wavefunction~\cite{lau1}
while the other FQHE states at
filling factors, $\nu=p/(2m p+1)$ ( p,m - integer) are 
readily understood in terms of the composite fermion (CF) theory.~\cite{jain}
The limit ($p \rightarrow \infty$) of such FQHE 
states~\cite{kamjg,orepjb}
corresponds to even-denominator filled states, $\nu=1/(2m)$ which are 
believed to be compressible Fermi liquid states qualitatively 
different from the 
FQHE states of the originating sequence.~\cite{hlr}
The basic understanding now is that the low-temperature phase of
fully spin-polarized electrons at filling fraction $\nu=1/(2m)$ is
a CF Fermi liquid phase.
A trial wave function for the CF Fermi liquid
ground state at $\nu=1/(2m)$ has been written down by Rezayi and 
Read~\cite{rezayi} and reads: 
\begin{equation}
\Psi_{Fermi} = \hat{P}_{LLL} \left[ det |e^{i \vec{k}_{\alpha} \vec{r}_i} | \
\Psi_{Bose} \right]  \ ,
\label{rrwf}
\end{equation}
where  $\Psi_{Bose}$ is the Bose Laughlin wave function for filling 
factor $\nu=1/(2m)$:
\begin{equation}
 \Psi_{Bose}=\prod_{i<j}^{N} (z_i-z_j)^{2m} \
 \exp{\left( -\sum_{j=1}^{N} \frac{|z_j|^2}{4 l_0^2} \right)} \ .
\label{bosewf}
\end{equation}
The Rezayi-Read (RR) CF Fermi wave function is obtained
after the product of the Bose Laughlin wave function with a 
Slater determinant of plane waves is
fully projected into the LLL by means of the projection 
operator, $\hat{P}_{LLL}$.
In the above expressions, $N$ is the number of electrons
that occupy the $N$ lowest-lying single-particle plane wave states
labeled by the momenta $\{ \vec{k}_{\alpha} \}$ consistent with an
ideal 2D spin-polarized Fermi gas,
$z_j=x_j+i y_j$ is the position coordinate for the j-th electron in complex
notation,
$l_0=\sqrt{\hbar/(eB)}$ is the magnetic length, $B$ is the
perpendicular magnetic field and $-e(e>0)$ is the electron's charge.

A straightforward motivation to study the Bose Laughlin 
wave function in relation to the true CF liquid state is justified
on the naive expectation that, because of the Bose statistics, the
Bose Laughlin energy may constitute a reasonable {\it lower energy bound } 
to the CF Fermi liquid energies for the given even-denominator filling factors.
While this is an expectation that we cannot prove exactly, 
an "a-posteriori" justfication may be provided if the numerical results are 
consistent with it (although we caution that this is not a definitive 
proof in any way).
The assumption that corresponding Bose Laughlin and CF liquid ground state 
energies should be reasonably different from  each other (given the very 
different nature of the wave functions) is more relevant to us in contrast 
to the question of whether the Bose Laughlin wave function is an exact 
lower energy bound of the true CF liquid energy.

The main result of this work is not to assert the relevance of
the Bose Laughlin wave function at filling factors $1/2$, $1/4$, and
$1/6$, but to report a surprising and unexpected finding, namely that 
the energy of the Bose Laughlin wave function at $\nu=1/4$ and $1/6$ 
(but not $1/2$) is practically identical to the energy of the 
corresponding true CF Fermi liquid state.

The model that we adopt consists of $N$ fully spin-polarized
electrons moving in a two-dimensional (2D) space subject to a strong
perpendicular magnetic field.
The electrons are embedded in a uniform  neutralizing
background of positive charge which is spread uniformly
within a finite disk of area $\Omega_N=\pi \, R_{N}^2$ and radius $R_N$.
The electrons can move freely all over the 2D space and are not constrained
to stay inside the disk.
The uniform density of the system is $\rho_0=\nu/(2 \pi l_0^2)$.
The radius of the disk is determined from the condition:
$\rho_0=N/\Omega_{N}$ which gives a finite disk radius
$R_N=l_0 \sqrt{2 N/\nu}$.
The kinetic energy per electron corrisponding to the
Bose Laughlin wave function is 
$\langle \hat{K} \rangle/N=\hbar \, \omega_c/2$, 
where $\omega_c$ is the cyclotron frequency.
Since the kinetic energy per electron is a mere constant the problem
reduces to calculate the expectation value of the potential
energy operator which consists of three terms:

\begin{equation}
\hat{V}=\hat{V}_{ee}+\hat{V}_{eb}+\hat{V}_{bb} \ , 
\label{pot_en}
\end{equation}
where $\hat{V}_{ee}, \hat{V}_{eb}$ and $\hat{V}_{bb}$ 
are, respectively, the electron-electron, electron-background and
background-background potential energy operators.
For a pure Coulomb interaction they are given by:
\begin{equation}
\hat{V}_{ee}=\sum_{i<j}^{N} \frac{e^2}{|\vec{r}_i-\vec{r}_j|} \ \ ; \ \
\hat{V}_{eb}=-\rho_0 \sum_{i=1}^{N} \int_{\Omega_N} d^2r \,
\frac{e^2}{|\vec{r}_i-\vec{r} |}  
   \ \ ; \ \
\hat{V}_{bb}=\frac{\rho_0^2}{2} \int_{\Omega_N} d^2r \int_{\Omega_N} 
d^2r^{\prime} \ \frac{e^2}{|\vec{r}-\vec{r}^{\ \prime}|} \ .
\label{potential}
\end{equation}
Monte Carlo (MC) simulations in disk geometry~\cite{morf,mcdisk} are used
to calculate the expectation value of the electron-electron
and electron-background interaction potentials,
$\langle \hat{V}_{ee} \rangle/N$ and
$\langle \hat{V}_{eb} \rangle/N$, respectively.
The background-background interaction energy can be
calculated exactly and is:
\begin{equation}
\frac{\langle \hat{V}_{bb} \rangle}{N}=
\frac{8}{3 \pi} \sqrt{\frac{\nu N}{2}} \frac{e^2}{l_0}  \ .
\label{bbintegral}
\end{equation}

In our MC simulations we adopt the wellknown Metropolis 
algorithm \cite{metropolis}.
In this algorithm, the expectation value of any 
operator is estimated by averaging its value over numerous 
electronic configurations of $\{\vec{r}_1, \ldots ,\vec{r}_N \}$ coordinates.
Our runs consist of 100,000 ``equilibration'' 
MC steps ( a step consists of attempts to move all the electrons one by one )
and up to
$2 \times 10^6$ ``averaging'' MC steps used to calculate the expectation
value of chosen operators.
The simulations are done for systems of
$N=4,16,36,64,100,144, 196, 256, 324$ and $400$ electrons.
The correlation energy per particle
in the thermodynamic limit ($N \rightarrow \infty$)
is obtained by fitting the finite-$N$ energies,
$\langle \hat{V} \rangle/N$ to a second-order
polynomial function as described  in Ref.~\cite{morf}.
The resulting polynomial fits are:
\begin{equation}
\left\{
\begin{array}{l}
\frac{\langle \hat{V} \rangle_{1/2}}{N}= 
    \left( -0.4841(5)+ \frac{0.1017(68)}{\sqrt{N}}-\frac{0.0280(07)}{N} \right)
    \frac{e^2}{l_0} \ ,  \\ 
\frac{\langle \hat{V} \rangle_{1/4}}{N}= 
    \left( -0.3614(03)+ \frac{0.0306(964)}{\sqrt{N}}-
                     \frac{0.0184(964)}{N} \right) \frac{e^2}{l_0} \ , \\ 
\frac{\langle \hat{V} \rangle_{1/6}}{N}= 
    \left( -0.3013(31)+ \frac{0.0133(206)}{\sqrt{N}}-
                 \frac{0.0151(46)}{N}  \right) \frac{e^2}{l_0}  \ .
\label{fit_std3}
\end{array}
\right.
\end{equation}

\begin{figure}[!t]
\onefigure[scale=0.5]{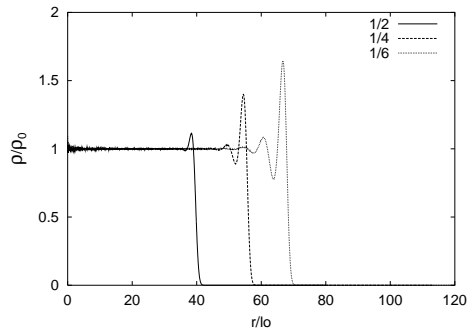}
\caption[]{One-body density function, $\rho(r)/\rho_0$, for the Bose
           Laughlin states $\nu=1/2,1/4$ and $1/6$ as a function of the
           distance $r/l_0$ from the center of the disk for
           systems with $N=400$ electrons.
            }
\label{rho_long_400}
\end{figure}

\begin{figure}
\onefigure[scale=0.5]{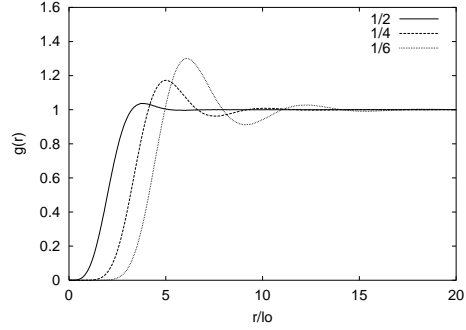}
\caption[]{Pair distribution function of the Bose Laughlin
           state at filling factor $\nu=1/2,1/4$ and $1/6$ 
            obtained after a MC simulation
            in disk geometry for $N=400$ electrons. }
\label{g_long_400}
\end{figure}

The finite-$N$ correlation energies per particle for the Bose Laughlin state
at $\nu=1/2,1/4$ and $1/6$ 
and their thermodynamic limit values are displayed in Table~\ref{tabstandard}.
 We estimated that the statistical uncertainty [which scales as 
 $1/\sqrt{MC steps}$] of the energy data reported in Table~\ref{tabstandard}
 is in the fifth digit after the decimal point which is rounded.
 Similarly, the statistical uncertainty of the extrapolated values is
 in the fifth digit after the decimal point.
\begin{table}[!ht]
\caption[]{Correlation energy per particle for the Bose Laughlin state
           at filling factors: $\nu=1/2,1/4$ and $1/6$. 
           Results obtained from a standard Monte Carlo
           simulation in disk geometry.
           Energies are in units of $e^2/l_0$.
           The statistical uncertainty in the last digit of energy is 
           shown in paranthesis. }
\label{tabstandard}
\begin{center}
\begin{tabular}{|c|c|c|c|}
\hline                                                  
N     & 1/2               & 1/4               & 1/6          \\ \hline
4     & -0.4402(8)        & -0.3506(9)        & -0.2984(7)   \\ \hline
16    & -0.4603(8)        & -0.3548(1)        & -0.2988(6)   \\ \hline
36    & -0.4679(7)        & -0.3568(2)        & -0.2995(6)   \\ \hline
64    & -0.4719(2)        & -0.3579(0)        & -0.2999(6)   \\ \hline
100   & -0.4743(2)        & -0.3585(7)        & -0.3002(1)   \\ \hline
144   & -0.4759(2)        & -0.3590(2)        & -0.3003(6)   \\ \hline
196   & -0.4770(5)        & -0.3593(2)        & -0.3004(7)   \\ \hline
256   & -0.4779(0)        & -0.3595(5)        & -0.3005(5)   \\ \hline
324   & -0.4785(3)        & -0.3597(2)        & -0.3006(0)   \\ \hline
400   & -0.4790(4)        & -0.3598(4)        & -0.3006(3)   \\ \hline
$\infty$  & -0.4841(5)    & -0.3614(03)       & -0.3013(31)  \\ \hline
\end{tabular}
\end{center}
\end{table}
We also computed the single-particle density function,
$\rho(\vec{r}) = \left\langle \sum_{i=1}^{N} 
\delta(\vec{r}-\vec{r}_i) \right\rangle$
and the pair distribution function which for a liquid with
uniform density $\rho_0$ is
$\rho_0 \ g(r)=\frac{1}{N} \left\langle \sum_{i=1}^{N} \sum_{j \neq i}^{N}
\delta (r-|\vec{r}_i-\vec{r}_j|) \right\rangle$.
The single-particle density of the Bose Laughlin state shows the
expected nonuniformity near the disk boundary (see Fig.\ \ref{rho_long_400}).  
Note that the non-uniformity near the disk edge always persists and 
increases as the filling factor decreases.  
The pair distribution function for $\nu=1/2,1/4$ and $1/6$ 
obtained after a MC simulation in disk geometry for $N=400$ electrons
is shown in Fig.\ref{g_long_400}.

We compare the Bose Laughlin energies in the thermodynamic limit 
as displayed in last row of
Table.~\ref{tabstandard} to respective values
of the LLL-projected spin-polarized CF Fermi
liquid state energies [Eq.(\ref{rrwf})] at same filling factors.
A search in the literature provides us with the following values
for the CF Fermi liquid energy at $\nu=1/2$:
$-0.466 \, e^2/l_0$ from Ref.~\cite{morfamb} and 
$-0.46557(6) \, e^2/l_0$ from Ref.~\cite{park}.
Both of these values are higher than the value
$-0.4841(50) \, \, e^2/l_0$ which corresponds to $\Psi_{Bose}$.
Ref.~\cite{morfamb} gives the energy:
$-0.3608 \, e^2/l_0$ for the $\nu=1/4$ state, a value that is
higher but very close to the energy,
$-0.3614(03) \, e^2/l_0$ corresponding to the $\Psi_{Bose}$ state.
We were unable to find data for the energy of the
$\Psi_{Fermi}$ state at $\nu=1/6$.
While there are more data for the {\it unprojected} version
of the CF Fermi liquid wave function obtained with different 
methods~\cite{chakraborty,orion_fhnc,orion_efhnc,orion_physe}, 
we found only few results for the {\it fully LLL projected} CF Fermi
liquid states at the filling factors considered in this work.
In order to have a larger pool of results to whom to compare the
$\nu=1/2, 1/4$ and $1/6$ energies we pursue a different path 
and resort to interpolation formulas that give the energy per particle,
$E(\nu)$ of the electronic liquid state as a function of the
LLL filling factor: $0 < \nu \leq 1$.
A crude empirical formula of this form has been suggested by
Laughlin (L)~\cite{lau1} and reads: 
\begin{equation}
E_{L}(\nu)=0.814 \, \sqrt{\nu} \, \left ( 0.230 \, \nu^{0.64} -1 \right ) \,
                \frac{e^2}{l_0} \ .
\label{elaughlin}
\end{equation}
Another formula proposed by Levesque, Weis, and 
MacDonald (LWM)~\cite{LWM} has the form:
\begin{equation}
E_{LWM}(\nu) =-0.782133 \, \sqrt{\nu} \, 
\left ( 1-0.211 \, \nu^{0.74} + 0.012 \, \nu^{1.7} \right ) \,
                \frac{e^2}{l_0} \ .
\label{elevesque}
\end{equation}
Both expressions, $E_L(\nu)$ and $E_{LWM}(\nu)$ are derived
by fitting the energy of Laughlin wave functions
and violate the particle-hole symmetry condition in the LLL.

A third interpolation formula which satisfies the 
particle-hole symmetry condition in the LLL and therefore 
provides a more accurate interpolation 
for the dependence of energy on the LLL filling factor is
given by Fano and Ortolani (FO)~\cite{fano} and has the form:

\begin{equation}
E_{FO}(\nu)= \left [-\sqrt{\frac{\pi}{8}} \, \nu-0.782133 \, \sqrt{\nu} \, 
             (1-\nu)^{3/2}  
                  +0.55 \, \nu \, (1-\nu)^2 
                       -0.463 \, \nu^{3/2} \, (1-\nu)^{5/2} \right]
                \frac{e^2}{l_0} \ .
\label{efano}
\end{equation}

We use the 
$E_{L}(\nu), E_{LWM}(\nu)$ and $E_{FO}(\nu)$ 
interpolation formulas
to generate more data for the energies at $\nu=1/2, 1/4$ and $1/6$
and compare the interpolated values 
to Bose Laughlin energies at same filling factors.
Such energies (in units of $e^2/l_0$ ) are given in Table~\ref{tabinf}.

\begin{table}[!t]
\caption[]{Energy per particle for the LLL states
           $\nu=1/2,1/4$ and $1/6$ obtained from interpolation
           formulas: Eq.(~\ref{elaughlin}), Eq.(~\ref{elevesque}) and
           Eq.(~\ref{efano})
           as compared with Bose Laughlin state energies.
           Energies are in units of $e^2/l_0$.
           The statistical uncertainty in the last digits of energy is 
           shown in paranthesis. 
}
\label{tabinf}
\begin{center}
\begin{tabular}{|c|c|c|c|}
\hline                                                  
$E(\nu)$           & 1/2         & 1/4         & 1/6               \\ \hline
$E_{L} $           & -0.490632   & -0.368452   & -0.308033          \\ \hline
$E_{LWM}$          & -0.485225   & -0.361930   & -0.301595         \\ \hline
$E_{FO} $          & -0.469049   & -0.361519   & -0.303660         \\ \hline
Ref.~\cite{morfamb}    
                   & -0.46600    & -0.36080    &  $\ldots$         \\ \hline
Ref.~\cite{park}    
                   & -0.46557(6) & $\ldots$    &  $\ldots$         \\ \hline
$E_{Bose}$         & -0.4841(50)   & -0.3614(03)  & -0.3013(31)   \\ \hline
\end{tabular}
\end{center}
\end{table}
The results in Table.~\ref{tabinf} and
the $\nu=1/2$ results from Ref.~\cite{morfamb,park} 
suggest that
the $E_L(\nu)$ and $E_{LWM}(\nu)$ 
formulas are too crude to describe the energy of the electronic liquid states
for the whole range of LLL filling factors.
Clearly the Bose Laughlin energy at $\nu=1/2$ is distinctly smaller 
than the CF Fermi liquid energy of Ref.~\cite{morfamb,park} 
and FO interpolation energy, as well.
The FO value at $\nu=1/2$ is close to the spin-polarized CF liquid
state value~\cite{park}, although we notice that it is closer
to the energy of the spin-unpolarized (singlet) CF state,
$-0.46953(7) \, e^2/l_0$ reported in Ref~\cite{park}.
This is not the case for the other two filling factors, 
$\nu=1/4$ and $\nu=1/6$, where all the interpolation energies
including the FO values are slightly lower than the Bose Laughlin
energies, though very close to them.
The more accurate polarized CF Fermi liquid
state energy~\cite{morfamb} 
($-0.3608 \, e^2/l_0$ at $\nu=1/4$) is larger than the Bose Laughlin energy, 
though it is important to note that the two values are so close that
they can be considered practically identical.

The fact that  interpolation energies are slightly lower than the Bose Laughlin
energies at $\nu=1/4$ and $1/6$ surely does not invalidate the possibility 
of the Bose Laughlin wave function being a lower energy bound to the 
true CF Fermi liquid energy since it is
well known~\cite{halperincusp} that the true energy at any filling
factor (excluding the Laughlin filling factors) is always higher than the  
approximate energy obtained from L, LWM and FO interpolation formulas
that are smooth functions of $\nu$ and should be handled with care. 
For electronic LLL Laughlin wave functions (Fermi or Bose) and a 
bare Coulomb
potential, the Haldane pseudopotential parameters, $V_{m}$ are known
to be all positive and decrease monotonically as the relative angular
momentum, $m$ increases.
In contrast to the electronic case, the Haldane pseudopotential
parameters (for both bare and screened Coulomb potentials) 
corresponding to the CF Fermi liquid states in Eq.(~\ref{rrwf})  
show a distinct non-monotonic behavior with some of them even becoming 
negative for large values of relative angular momentum.~\cite{goerbig}
In the CF language, a state with filling factor, $\nu=p/(2m p+1)$ 
represents a CF system with $p$ filled CF LL-s which weakly interact with
each other by some residual interaction.
In such a case, the CF Haldane pseudopotentials (both small $p$ and large $p$)
show a qualitatively different behavior from the electronic LLL
Laughlin pseudopotentials.
Such a qualitatively different behavior is likely associated with the 
weakly interacting nature of the 
CF-s as compared to pure electrons.~\cite{wojs}
Because of the different nature of the two liquid wave functions, 
[Eq.(~\ref{rrwf}) and Eq.(~\ref{bosewf})] and their respective
Haldane pseudopotentials, it was unexpected to find that
the Bose Laughlin and the true CF Fermi liquid energies are practically 
identical for filling factor $\nu=1/4$ and $1/6$, while 
clearly distinct for $\nu=1/2$.

In summary, we used MC simulations to calculate the ground
state properties of Bose Laughlin wavefunctions at compressible filling
fractions $\nu=1/2, 1/4$ and $1/6$. The calculated ground state energies are
compared with results for spin-polarized CF wavefunctions for these 
filling factors.
We found that the Bose Laughlin wavefunction ground state energy is 
lower than the CF Fermi liquid energy for $\nu=1/2$, whereas for other 
filling factors, $\nu=1/4$ and $1/6$ we find the surprising result that 
the Bose Laughlin ground state energy is practically identical to the energy 
of the true CF liquid state as calculated from different studies and from 
estimates obtained from various interpolation schemes.
Because Bose Laughlin and true CF liquid energies are 
very different at $\nu=1/2$
we naively expected to see a similar tendency apply to the other two 
even-denominator-filled CF liquid states, $\nu=1/4$ and $1/6$, however 
unexpectedly we found out that the Bose Laughlin and true CF liquid
energies are practically identical at filling factors $1/4$ and $1/6$.
The physical significance and robustness of this finding is not yet clear
to us and further work is needed to address this issue.

\acknowledgments
This research was supported by the U.S. D.O.E. (Grant No. DE-FG52-05NA27036)
and by the Office of the Vice-President
for Research and Development of Prairie View A\&M University through
a 2003-2004 Research Enhancement Program grant.


\end{document}